\documentstyle[epsf,floats,tighten,preprint,eqsecnum,prd,aps]{revtex}

\def\Lin{{L_{\text{in}}}}
\def\half{{1\over 2}}
\def\Tr{\text{Tr}\,}
\def\Trone{\text{Tr}_1\,}

\def\<{\langle }
\def\>{\rangle }
\def\vac{^{\text{vac}}}
\def\bx{{\bf x}}
\def\bz{{\bf z}}
\def\Px{{\bf P}_x}
\def\Pz{{\bf P}_z}
\def\Nout{{N_{\text{out}}}}
\def\Nin{{N_{\text{in}}}}
\def\Ninn{{N_{\text{in}}+1}}
\def\Nnorm{{N_{\text{norm}}}}
\def\Tunk{T_{\text{unk}}}
\def\pl{_{\text{pl}}}
\def\xalpha{{\bf x}^\alpha}
\def\rb{^{\text{rb}}}

\begin{document}

\title{Entropy of Localized States and Black Hole Evaporation\thanks
{This work is supported in part by funds provided by the U.S.
Department of Energy (D.O.E.) under cooperative 
research agreement \#DF-FC02-94ER40818.}}

\author{Ken D.\ Olum\footnote{Email address: {\tt kdo@ctp.mit.edu}}}

\address{Center for Theoretical Physics \\
Laboratory for Nuclear Science \\
and Department of Physics \\
Massachusetts Institute of Technology \\
Cambridge, Massachusetts 02139}

\date{MIT-CTP-2580, hep-th/9610086, November 1996}

\maketitle

\begin{abstract}%
We call a state ``vacuum-bounded'' if every measurement performed
outside a specified interior region gives the same result as in the
vacuum.  We compute the maximum entropy of a vacuum-bounded state with
a given energy for a one-dimensional model, with the aid of numerical
calculations on a lattice.  The maximum entropy is larger than it
would be for rigid wall boundary conditions by an amount $\delta S$,
which for large energies is $\alt {1\over 6} \ln (\Lin T)$, where
$\Lin$ is the length of the interior region.  Assuming that the state
resulting from the evaporation of a black hole is similar to a
vacuum-bounded state, and that the similarity between vacuum-bounded
and rigid-wall-bounded problems extends from 1 to 3 dimensions, we
apply these results to the black hole information paradox.  We
conclude that large amounts of information cannot be emitted in the
final explosion of a black hole.
\end{abstract}

\vspace*{\fill}
\begin{center}
Submitted to {\it Phys. Rev. D.}
\end{center}


\narrowtext


\section{Introduction}

\subsection{Background}

Since the discovery of black hole radiation by Hawking
\cite{hawking:orig}, the fate of information falling into a black hole
has been a mystery.  (See
\cite{page:review,preskill:review,banks:review} for reviews.)  If
Hawking's semiclassical calculation is correct, then the outgoing
radiation is purely thermal and the outgoing photons are uncorrelated
to each other and to the matter which formed the black hole.  If the
evaporation is complete, and if the thermal nature of the radiation
persists throughout the evaporation, then the original information is
lost.  Thus if the black hole is formed from a quantum-mechanically
pure state, there will nevertheless be a mixed state after the
evaporation is complete.  This is the position held by Hawking
(e.g. see \cite{hawking:virtual}), but it violates CPT and may lead to
difficulties with energy conservation and causality
\cite{page:review,giddings:infoloss,banks:pure-mixed,%
strominger:unitary}.

If information is not lost in black hole evaporation, there are
several possibilities. One is that the black hole does not evaporate
completely, but instead produces one or more Planck-scale remnants
(e.g. see \cite{banks:review}).  Another possibility is that the
information disappears into a baby universe \cite{strominger:baby}.
In this scenario the quantum-mechanical pure state is preserved, but
parts of it are inaccessible to observation.  It is also possible that
the radiation is not really thermal, even at early times, because of a
complementarity principle \cite{susskind:strings,verlinde:complement}
or the inapplicability of the semiclassical approach
\cite{esko:nosemiclass,bose:nosemiclass,casher:nosemiclass}, and thus
that the information is encoded in subtle correlations in the
radiation.  In this case the black hole could act like a normal object
with the entropy describing internal degrees of freedom.  Some results
from string theory \cite{strominger:string,esko:string} tend to
confirm this view.

Even if the radiation is thermal and uncorrelated during most of the
evaporation, there is no reason to believe that it remains thermal
near the endpoint of the evaporation.  The late-time radiation is
presumably governed by an unknown theory of quantum gravity, and may
well have correlations to the radiation emitted earlier.\footnote{This
is conceivable because at early times information in the outgoing
radiation can be correlated with information in the ingoing
negative-energy flux.}  However, it is generally believed that
late-time radiation cannot resolve the information paradox
\cite{aharonov:orig,preskill:review,banks:review}.  The argument goes
as follows: While the black hole is large, it is presumably radiating
high-entropy thermal radiation.  If the final explosion is to restore
a pure state, it must radiate as much entropy\footnote{Here and
throughout this paper, ``entropy'' means fine-grained
quantum-mechanical entropy.}  as was radiated in earlier times.
However, by the time the black hole reaches the point where unknown
physics could come into play, there is little energy remaining.  To
radiate a lot of information with little energy requires a long period
of time, and thus the ``final explosion'' looks more like a long-lived
remnant.

However, Wilczek \cite{wilczek:mirror} argues from a moving mirror
model that a state with high entropy can nevertheless be purified
with arbitrarily low energy cost.  In certain ways, his model looks
more like a remnant theory than a complete-evaporation theory, but it
still appears to cast some doubt on the standard argument above.

In any case, this argument requires bounding the entropy that can be
contained in a particular region with a fixed energy.  In the case of
a region with reflecting walls, this is the question of finding the
thermal state of quantum fields in a box.  For a spherical box and a
particular field theory the problem is easily solved.  But with a
region of complex shape, or where one wishes to make a statement
intended to apply to all field theories, the situation is more
complicated.  Bekenstein \cite{bekenstein:talk,bekenstein:argue}
argues that such a universal bound exists, although Unruh and Wald
\cite{unruh-wald:argue} disagree.

Here we take a different approach.  We consider only a single scalar
field, but we use a weaker and, we hope, more physical condition on
the results of the black hole evaporation.  In the end, our results
still support the claim that late-time radiation cannot restore the
purity of the state of an evaporating black hole.

\subsection{The ``vacuum-bounded state''}

We start by considering a black hole formed from a pure
quantum-mechanical state of incoming matter.  To avoid any possible
complications of quantum gravity theory, we will look at the state
produced after the black hole has completely evaporated
\cite{preskill:review}.  Gravity should play no significant role in
this state, since the energy density state should be small
everywhere.\footnote{If instead there are Planck-scale concentrations
of energy, then we would have a remnant theory, a possibility we are
explicitly not considering here.}  We can describe the final state as
follows: At large distances from the position of the black hole (taken
as the origin) there is outgoing Hawking radiation, which we are
assuming to be thermal and uncorrelated to anything.  Within some
distance $R$ of the origin, there is some state of ordinary quantum
fields that could have correlations with the radiation emitted
earlier.  The distance $R$ is the distance to which such information
might have propagated after unknown physics came into play.  Let us
assume that Hawking's semi-classical calculation is good up to an
energy scale $\Tunk$.  This temperature is reached when the black hole's
mass is\footnote{We are working with units in which $c = G = \hbar =
k_B = 1$} $M_0 = 1/(8\pi \Tunk)$.  If, after this, the rate of
evaporation continues to match the Hawking calculation,\footnote{As
opposed, for example, to slowing to nothing and leaving a remnant.}
the black hole will evaporate in time $t\sim {10^4 M^3/g} \sim {1/(g
\Tunk^3)}$,
 where $g$ is the effective number of degrees
of freedom in the particles that can be radiated.  (See
\cite{page:radiation}.)  So there is a sphere of radius
\begin{equation}
R \sim {1\over g \Tunk^3}
\end{equation}
 which contains total energy
\begin{equation}
E_0 = {1\over 8\pi \Tunk}
\end{equation}
 in which the information could be
contained.
\label{sec:realistic-er}
Taking, for example, 
$\Tunk = 10^{15} \text{Gev} \sim 10^{-4} m\pl$
and $g\sim 100$ we get\footnote{Another possibility is that $g$
diverges as $T\rightarrow m\pl$.  In this case the information can be
radiated in a small number of particles of about the Planck mass,
chosen from an infinite spectrum of such particles.  This is
effectively a remnant theory.}
\begin{mathletters}
\label{eqn:goshwownumbers}
\begin{eqnarray}
R & \sim & 10^{10}l\pl \sim 10^{-23}\text{cm}\\
E_0 & \sim & 10^3 m\pl \sim 10^{-2}\text{g} \sim 10^{19}\text{erg}\,.
\end{eqnarray}
\end{mathletters}

Now we would like to answer the following question: How much entropy
can be contained in a spherical region of radius $R$ with a total
energy of $E_0$?  To answer this question we have to specify what we
mean by ``contained in a region''.  As mentioned earlier, if we ask
how much entropy can be contained in a spherical box of radius
$R$ with perfectly reflecting walls, the question can be easily
answered.  However, the system with the box is not so closely akin to
the system under discussion.  For instance, inserting the reflecting
walls into the system produces a divergent increase in the
ground-state energy of the system.  Furthermore, if we started with
the vacuum in the whole system, and then introduced a spherical wall,
we would produce a divergent geometric
entropy\cite{srednicki:geom-ent,callan:geom-ent}.  A better
description of our system is simply that it has thermal radiation
outside radius $R$, and an unknown state of the quantum fields inside
radius $R$, but no barrier or boundary at $R$.

To study such systems we will assume that the difference between the
external Hawking radiation and an external vacuum is not important to
considerations of entropy.\footnote{If this approximation is bad we can
always increase $R$ until the Hawking radiation outside $R$ has very
low temperature.}  We will study systems that have an arbitrary
state inside $R$ but the vacuum outside $R$.  To make this precise we
will specify the problem as follows:

\begin{quotation}
Let a ``vacuum-bounded state'' be a generalized state (i.e.\ density
matrix) for which every operator composed of fields at points outside
a specified interior region has the same expectation value as in the
vacuum.  What is the maximum entropy of such a state whose interior
region is a sphere of radius $R$ and whose total average
energy\footnote{We cannot specify that any measurement of the energy
must give a particular value.  Such a state is necessarily static and
thus cannot represent outgoing radiation.} is given by $\<H\> = E_0$?
\end{quotation}

We expect to find that the answer to this question is similar to that
of a box of radius $R$ with reflecting walls, with some small
correction.

\section{General considerations}
We will first consider a general system divided into an inside region
(``part 1'') and an outside region (``part 2'').  We will say that a
generalized state (i.e.\ density 
matrix) is ``localized to the inside'' or ``obeys the vacuum-bounded
condition'' if any measurement performed on the outside field
operators in this state yields the same result as in the vacuum, i.e.\
if
\begin{equation}\label{eqn:vac-cond}
\Tr \rho O_2 = \Tr \rho\vac O_2 = \<0|O_2|0\>
\end{equation}
for every operator $O_2$ that is constructed out of field operators in
the outside region.

In the language of density matrices, we can write $\rho_2 =
\Trone\rho$, where $\rho$ is the overall density matrix describing our
system and $\Trone$ means to trace over all the ``inside'' variables.
Then $\rho_2$ is the reduced density matrix describing only the
outside variables.  Eq.\ (\ref{eqn:vac-cond}) then is
equivalent to 
\begin{equation}
\rho_2 = \rho_2\vac \equiv \Trone|0\>\<0|\,,
\end{equation}
where $|0\>$ denotes the ground state.

\subsection{Stationary points of $S$}
The present problem is a particular case of the following general
problem: Find a density matrix $\rho$ (i.e.\ a
positive-semidefinite Hermitian matrix with unit trace) which maximizes
$S = -\Tr\rho\ln\rho$ subject to a set of constraints $\Tr\rho C_\alpha
= V_\alpha$.  To analyze this problem we will look at the change in
$S(\rho')$ when $\rho$ is varied via $\rho' = \rho + t\delta\rho$.
The naive result holds for the first derivative,
\begin{equation}
{dS\over dt} = -\Tr\left(\delta\rho\ln\rho+\delta\rho\right)\,.
\end{equation}
We will only consider variations of $\rho$ which leave $\rho$ normalized with
$\Tr\rho = 1$, which means that $\Tr\delta\rho = 0$ and thus
\begin{equation}
{dS\over dt} = -\Tr\delta\rho\ln\rho\,.
\end{equation}

We show in appendix \ref{app:convex} that entropy is always strictly
(downward) concave \cite{lieb:convex} which means that
\begin{equation}
{d^2S\over dt^2} < 0
\end{equation}
 for any $\delta\rho$.
Thus any point at which $S$ is stationary is a local maximum.
Furthermore, we can show that there can be at most one such maximum,
as follows.

\label{sec:unique}
Let ${\cal P}$ be the space of all Hermitian matrices such that $\Tr\rho = 1$
and $\Tr\rho C_\alpha = V_\alpha$, and let ${\cal P}_+$ be the subspace
of ${\cal P}$ where $\rho$ is positive-semidefinite.
First we note that ${\cal P}_+$ is convex:  Let $\rho_1$ and $\rho_2$
be elements of ${\cal P}_+$
and let $\rho(t) = t\rho_1 + (1-t)\rho_2, t \in
[0,1]$.  The matrix $\rho(t)$ is clearly Hermitian, properly
normalized, and satisfies the constraints.  It is also positive
semidefinite, since for any $\bx$, $\bx\cdot\rho(t)\bx = t(\bx\cdot\rho_1\bx)
+ (1-t)(\bx\cdot\rho_2\bx) \ge 0$.  Thus $\rho(t)$ is an admissible
density matrix as desired.

Since $S$ is a strictly concave function defined on a convex space, it
cannot have more than one stationary point.  For if we had two
stationary points we could vary $\rho$ along a line between them, and
since $d^2S/dt^2 < 0$ we could not have $dS/dt=0$ at both.

If there is any $\rho$ which achieves the maximum value of $S$, it
must be a stationary point of
$S$.  The only conceivable alternative would be for $\rho$ to lie on
the edge of ${\cal P}_+$ in ${\cal P}$, but that is excluded by the
following argument:  Suppose $\rho$ is on the edge of ${\cal P}_+$ in
${\cal P}$.  Then there is some direction $\delta\rho$ such that
$\rho' \equiv \rho + t\delta\rho$ is positive-semidefinite for arbitrarily
small positive $t$ but not for negative $t$ arbitrarily close to zero.
That can happen only if $\rho$ has arbitrarily small or zero
eigenvalues $\rho_i$ which become negative as $t$ becomes negative.
For this to happen these eigenvalues must increase from arbitrarily
small values as $t$ becomes positive.  As $t\rightarrow 0$ these
eigenvalues dominate everything else in $dS/dt = \sum_i \left(-
\ln\rho_i\right)$, so $dS/dt \rightarrow +\infty$.  Thus for sufficiently small
but positive $t$, $S(\rho') > S(\rho)$, so $\rho$ does not maximize
$S$.

Whether a maximum-entropy point exists at all depends on the
constraints $C_\alpha$.  First suppose there is a constraint on the
energy $\Tr \rho H = E_0$ and no other constraints.  In a normal
system there will be a state with the maximum entropy.\footnote{But
note that if the density of states rises exponentially or faster
with energy, then the entropy can be unbounded.}  Now suppose that there are
additional constraints.  They further restrict the space of allowable
$\rho$ and thus can only reduce the achievable values of $S$, so there
is at least a supremum of possible values of $S$ in the constrained
system.  We believe but have not been able to prove that in our
problem there will be a $\rho$ which achieves the maximum value of
$S$.  In the numerical work described in Sec.\ \ref{sec:numerical} we
have always succeeded in finding a solution.

\subsection{The form of $\rho$}

We must find the unique state $\rho$ which satisfies our
constraints and which gives $\Tr\delta\rho\ln\rho = 0$ for any
$\delta\rho$ which maintains the constraints.
This means that we are concerned with $\delta\rho$ such that
\begin{mathletters}
\label{eqn:drho}
\begin{eqnarray}
\Tr\delta\rho & = & 0 \\
\Tr\delta\rho C_\alpha & = & 0
\end{eqnarray}%
\end{mathletters}%
for all $\alpha$.

Now we can treat the space of Hermitian operators as an inner product
space with
$(A, B) \equiv \Tr A B$.  In this language, Eqs.\ (\ref{eqn:drho}) imply
that the possible $\delta\rho$ span
the space orthogonal to the identity and to each of the $C_\alpha$.
Since $\ln\rho$ must be orthogonal to all the $\delta\rho$, it must
be composed only of
$I$ and the $C_\alpha$, so we can write
\begin{equation}
\ln\rho = {\text{const}} + \sum_\alpha f_\alpha C_\alpha
\end{equation}
for some coefficients $f_\alpha$.  Since $\Tr\rho = 1$, we
can write
\begin{equation}
\rho = {e^{\sum f_\alpha C_\alpha} \over \Tr e^{\sum f_\alpha C_\alpha}}\,.
\end{equation}
Our goal is now to determine the coefficients $f_\alpha$ so that the
constraints are satisfied.

We can define a grand partition function,
\begin{equation}
Q=\Tr e^{\sum f_\alpha C_\alpha}\,.
\end{equation}%
Its derivatives are
\begin{equation}
{dQ\over df_\alpha} = \Tr C_\alpha e^{\sum f_\alpha C_\alpha}\,,
\end{equation}
so
\begin{equation}
\<C_\alpha\> = {d\over df_\alpha}\ln Q\,.
\end{equation}
  We have the usual
thermodynamic formula for the entropy,
\begin{equation}
S = -\<\ln\rho\> = \ln Q - \sum f_\alpha \<C_\alpha\>\,.
\end{equation}
Differentiating this we find
\begin{equation}
{dS\over df_\alpha} = - \sum f_\beta {d\<C_\beta\>\over df_\alpha}\,.
\end{equation}

Now we specialize to the case where one of the constraints is
just the Hamiltonian.  The corresponding coefficient is written
$-\beta$, and we have
\begin{equation}
\rho = {1\over Q}e^{-\beta H + \sum f_\alpha C_\alpha}\,.
\end{equation}
If we vary the coefficients in such a way that $\<H\> = E$ changes but
the other $\<C_\alpha\> = V_\alpha$ remain fixed we see that
\begin{equation}
{dS = - \sum f_\alpha d\<C_\alpha\> = \beta dE}\,.
\end{equation}
 Thus $\beta =
dS/dE$ and so the coefficient $\beta$ has the usual interpretation as
the inverse temperature.

\section{One scalar field, one dimension.}
Now we will restrict ourselves to a theory consisting only of
gravity and one massless scalar field.  Presumably such a system has enough
richness to contain the usual black hole information paradox, so
nothing important is lost by making this reduction.  We will also
begin here by working in one dimension.  We will put our entire system
in a box of length $L$ and require that all deviations from the vacuum
are in the region from $0$ to $\Lin$.  Later we will let the
overall box size $L$ go to infinity while $\Lin$ remains fixed.  The
inside region will be $[0, \Lin]$ and the outside
region will be $[\Lin, L]$.  We will use the usual
scalar field Hamiltonian, which in classical form is
\begin{equation}
H = \half \int_0^L \left[\pi(x)^2 
 + \left({d\phi\over dx}\right)^2\right] dx\,.
\end{equation}

\section{The solution density matrix is Gaussian}
In our problem, the constraints are the energy bound,
\begin{equation}
\Tr\rho H = E_0\,,
\end{equation}
and the ``vacuum-bounded'' condition,
\begin{equation}
\Tr\rho O_2^\alpha = \<0|O_2^\alpha|0\>\,,
\end{equation}
where $O_2^\alpha$ is any operator which is constructed out of the
fields $\phi(x)$ and $\pi(x)$ in the outside region.  The solution
must have the form
\begin{equation}
\rho \propto e^{-\beta H + \sum f_\alpha O_2^\alpha}\,.
\end{equation}
We now show that $f_\alpha$ is nonzero only for those
operators $O_2^\alpha$ which are quadratic in the fields.

Suppose that we wanted to solve a different problem in which we cared
only about the constraints involving the quadratic operators.  We would
have the energy bound and the constraints
\begin{equation}
\Tr\rho Q_2^a = \<0|Q_2^a|0\>\,,
\end{equation}
where $Q_2^a$ runs only over quadratic operators $\phi(x)\phi(y)$ and
$\pi(x)\pi(y)$.  (The operators $\phi(x)\pi(y)$ vanish automatically
by symmetry under $\phi \rightarrow -\phi$.)  The solution to this
problem has the form
\begin{equation}
\rho' \propto e^{-\beta H + \sum f_a Q_2^\alpha}\,.
\end{equation}

Now $\rho'$ is a Gaussian operator, i.e.\ $\<\phi(\cdot)|\rho'|\phi'(\cdot)\>$
is a Gaussian functional of the values of $\phi$ and $\phi'$.  Let
$\rho'_2 = \Trone \rho'$.  The trace is just a set of Gaussian
integrals, which means that the resulting $\rho'_2$ is also a
Gaussian.  Because $H$ is quadratic, the vacuum $\rho\vac = |0\>\<0|$ is
Gaussian, and so is its
trace $\rho\vac_2 = \Trone \rho\vac$.  Now by construction we have
$\Tr\rho'_2 \phi(x)\phi(y) = \Tr\rho\vac_2 \phi(x)\phi(y)$ and
$\Tr\rho'_2 \pi(x)\pi(y) = \Tr\rho\vac_2 \pi(x)\pi(y)$.  These
conditions are sufficient to fix the coefficients in the Gaussian
$\rho'_2$, and thus to show that in fact $\rho'_2$ and $\rho\vac_2$ are the
same Gaussian, i.e.\ that $\Trone\rho' = \Trone\rho\vac$.

Thus $\rho'$ satisfies all the constraints of the original problem.
Since only one $\rho$ can have these properties it follows that
$\rho = \rho'$ and thus that Gaussian solution $\rho'$ is the correct
solution to the original problem.



\section{The discrete case}
We now approximate the continuum by a one-dim\-en\-sional lattice of coupled
oscillators, with a classical Hamiltonian
\begin{equation}\label{eqn:hdisc}
H = \half\left(\Px \cdot \Px + \bx \cdot K \bx\right)\,.
\end{equation}
The simple kinetic term in Eq.\ (\ref{eqn:hdisc}) corresponds to
choosing oscillators of unit mass, regardless of how densely they are
packed.  In terms of scalar-field variables this means that $x_i =
\sqrt{a}\phi_i$ and $P_i = \pi_i/\sqrt{a}$ where $a = L/(N+1)$
is the lattice spacing, $\phi_i$ is the average of
$\phi(x)$ over an interval of length $a$, and $\pi_i$ is the
total momentum $\pi(x)$ in the interval.

The matrix $K$ gives the couplings between the oscillators and
represents the $d\phi/dx$ term in the scalar field Hamiltonian.  To
approximate the continuum with the zero-field boundary condition we
will imagine that we have $N$ oscillators located at the points
$1/N+1\ldots N/N+1$.  and the end oscillators are coupled to
fixed-zero oscillators at $0$ and $1$.  Then

\begin{equation}
K = \left(\matrix{2g & -g & 0 & 0 & \hspace{5pt}\cdot\hspace{5pt} \cr
                    -g & 2g & -g & 0 & \cdot \cr
                    0 & -g & \cdot & \cdot & \cdot \cr
                    0 & 0 & \cdot & \cdot & \cdot \cr
                    \cdot & \cdot & \cdot & \cdot & \cdot \cr
                    }\right)
\end{equation}
where $g = (N+1)^2/L^2$.

We are trying to satisfy the constraints
\begin{mathletters}
\begin{eqnarray}
\Tr\rho H & = & E_0\\
\Tr\rho x_i x_j & = & \<0|x_i x_j|0\>\\
\Tr\rho P_i P_j & = & \<0|P_i P_j|0\>\,,
\end{eqnarray}%
\end{mathletters}%
where $i$ and $j$ run over the oscillators which represent 
the outside region. We have shown that the solution will have the
form
\begin{equation}
\rho \propto e^{-\beta H + f^{(x)}_{\mu\nu} x_\mu x_\nu
   + f^{(P)}_{\mu\nu} P_\mu P_\nu}\,.
\end{equation}
We can write this in a more familiar form as
\begin{equation}
\rho \propto e^{-\beta H'}\,,
\end{equation}
where $H'$ is a fictitious Hamiltonian for these oscillators,
\begin{equation}
H' = \half\left(\Px\cdot T'\Px + \bx\cdot K'\bx\right)
\end{equation}
with
\begin{equation}
T' = \left(\begin{array}{c|c}
	I & 0\\ \hline
	0 & T'_{22}\\
	\end{array}\right)
\end{equation}
and
{\renewcommand\arraystretch{1.3}
\begin{equation}
K' = \left(\begin{array}{c|c}
	K_{11} & K_{12}\\ \hline
	K_{12}^T & K'_{22}\\
	\end{array}\right)\,.
\end{equation}
}%
Here $K_{11}$ and $K_{12}$ are the sections of
the original coupling matrix $K$, and $T'_{22}$ and $K'_{22}$ and
$\beta$ can be adjusted in an attempt to meet the necessary
conditions.  This gives us one number, $\beta$, and two symmetric
$\Nout$-by-$\Nout$ matrices, $K'_{22}$ and $T'_{22}$, that we can
adjust.  The constraints involve one scalar constraint, for $H$, and
two $\Nout$-by-$\Nout$ symmetric matrices of constraints, for $x_i
x_j$ and $P_i P_j$.  There are equal numbers of
equations to satisfy and free parameters to adjust, and so, if we are
lucky, we will be able to find a solution.  If we do find a solution,
we know it is unique from the arguments of Sec.\ \ref{sec:unique}.

\subsection{Computing the expectation values}

To actually solve these equations we will need to compute the
expectation values of $x_i x_j$ and $P_i P_j$ given the density matrix
$\rho \propto \exp\{-\beta H'\}$.  We can compute them in the usual
way as derivatives of the partition function
\begin{equation}
Q = \Tr e^{-\beta\cdot\half\left(K'_{\mu\nu}x_\mu x_\nu
+ T'_{\mu\nu}P_\mu P_\nu\right)}\,,
\end{equation}
so
\begin{mathletters}
\begin{eqnarray}
\<x_i x_j\> & = & -{1\over\beta}\left({d\over dK'_{ij}}+{d\over dK'_{ji}}\right)\ln Q \\
\<P_i P_j\> & = & -{1\over\beta}\left({d\over dT'_{ij}}+{d\over dT'_{ji}}\right)\ln Q\,.
\end{eqnarray}
\end{mathletters}

To compute the partition function we will first find the normal modes of
the fictitious classical problem with this Hamiltonian, and then treat
these as independent oscillators which we will quantize.
The Hamiltonian $H'$ gives rise to the equations of motion
\begin{equation}
{d^2 x_\mu\over dt^2} = - T'_{\mu\nu} K'_{\nu\lambda} x_\lambda\,,
\end{equation}
so we look for eigenvectors $\xalpha$ that satisfy
\begin{equation}
T' K' \xalpha = \omega_\alpha^2 \xalpha\,.
\end{equation}
The eigenvectors will be complete, so that we can define new
coordinates $z_\alpha$ via $\bx = \sum z_\alpha \xalpha$, which
will then obey the equations of motion
\begin{equation}
{d^2z_\alpha\over dt^2} = - \omega_\alpha^2 z_\alpha\,.
\end{equation}
We can choose the norms of the eigenvectors so that
$\xalpha\cdot K'\bx^\beta = \omega_\alpha^2 \delta_{\alpha\beta}$
and group the eigenvectors into a matrix $V$ via $V_{\mu\alpha} = x^\alpha_\mu$.
Then we will find that
$T'_{\alpha\beta} = x^\alpha_\mu x^\beta_\mu$, or
\begin{equation}
K' = {V^{-1}}^T \Omega^2 V^{-1} \qquad\text{and}\qquad T' = V V^T 
\end{equation}
where $\Omega_{\alpha\beta} = \omega_\alpha \delta_{\alpha\beta}$.
We can then substitute
\begin{equation}
\bx = V \bz\qquad \Px = {V^{-1}}^T \Pz
\end{equation}
into $H'$ to get
\begin{eqnarray}
H' &=& \half\left(\Pz\cdot\Pz+\bz\cdot\Omega^2\bz\right)\nonumber\\
&=& \half\sum_i(P_\alpha^2+\Omega_\alpha^2z_\alpha^2) 
\equiv \sum_\alpha H'_\alpha\,.
\end{eqnarray}

This is the Hamiltonian for a set of (fictitious) uncoupled
oscillators with frequencies $\omega_i$, and thus
$Q = \Tr e^{-\beta H'}$ will be the product of partition functions
$Q_\alpha = \Tr e^{-\beta H'_\alpha}$ for the individual oscillators.  The
partition function of a single oscillator is easily found to be
\begin{equation}
Q_i = {e^{-\beta\omega_\alpha/2}\over 1-e^{-\beta\omega_\alpha}}
= {1\over 2} \text{csch}\,{\beta\over 2}\omega
\end{equation}
and thus
\begin{equation}
\ln Q = \sum \ln Q_\alpha = - \sum \ln \left(2 \sinh{\beta\over 2}\omega_\alpha\right)\,.
\end{equation}
Differentiating gives
\begin{equation}
{d\over d\omega_\alpha} \ln Q = -{\beta\over 2}\coth{\beta\over 2}\omega_\alpha\,.
\end{equation}
Letting $\Lambda_\alpha = \omega_\alpha^2$, this becomes
\begin{equation}
{d\over d\Lambda_\alpha}\ln Q = -{\beta\over 4\omega_\alpha}\coth{\beta\over 2}\omega_\alpha\,.
\end{equation}
We can compute the effect of a change to $K'_{ij}$ or $T'_{ij}$ on the
eigenvalues,
\begin{equation}
{d\Lambda_\alpha\over dK'_{ij}} = V_{i\alpha} V_{j\alpha}
\end{equation}
and
\begin{equation}
{d\Lambda_\alpha\over dT'_{ij}} = V^{-1}_{\alpha i} V^{-1}_{\alpha j} \Lambda_\alpha\,.
\end{equation}
Finally,
\begin{mathletters}
\label{eqn:nonlineartosolve}
\begin{eqnarray}\label{eqn:xxcorrelator}
\<x_i x_j\> = \sum_\alpha{1\over 2\omega_\alpha}V_{i\alpha} V_{j\alpha}\coth{\beta\over2}\omega_\alpha\\
\<P_i P_j\> = \sum_\alpha{\omega_\alpha\over 2}V^{-1}_{\alpha i} V^{-1}_{\alpha j}
\coth{\beta\over 2}\omega_\alpha\,.
\end{eqnarray}
\end{mathletters}

These expressions must have the same values when $V$ and the $\omega$
are the eigenvectors and frequencies of $T'K'$ as they have in the
vacuum, i.e.\ when $V$ and $\omega$ are the normal modes and
frequencies of the vacuum and $\beta=\infty$.

\section{The form of the normal modes}

We are trying to find the normal mode eigenvectors $\bx^\alpha$ and
frequencies $\omega_\alpha$ that will satisfy our constraints.  If we
take the $\omega_\alpha$ as given, we can learn about the
$\bx^\alpha$ as follows:
The vectors $\xalpha$ satisfy the equation 
\begin{equation}
T' K' \xalpha = \omega_\alpha^2 \xalpha
\end{equation}
with
\begin{equation}
T' = \left(\begin{array}{c|c}
	I & 0\\ \hline
	0 & T'_{22}\\
	\end{array}\right)
\end{equation}
and


{\arraycolsep 0pt
\begin{equation}
K' = \left(\begin{array}{c|c}
K_{11} & \begin{array}{cc}
	 & \hspace{4pt} 0 \hspace{4pt} \\
	 \cline{1-1}
	 \mbox{\hspace{2pt} \small $-g$ \hspace{1pt}} & \vline\hfill\end{array} \\
\hline
\begin{array}{cc}
\hfill\vline & \mbox{\hspace{2pt} \small $-g$ \hspace{1pt}} \\
\cline{2-2}
\hspace{4pt} 0 \hspace{4pt} & \end{array} & K'_{22} \end{array}\right)\,.
\end{equation}
}

Let us introduce the convention that Latin letters from the start of
the alphabet, $a,b,c,\ldots$ range over only ``inside'' oscillator
indices, Latin letters from the middle of the alphabet,
$i,j,k,\ldots$ range over only ``outside'' oscillator indices, and
Greek letters range over all indices.
Writing the eigenvalue equation out in components,
\begin{equation}
T'_{\mu\nu} K'_{\nu\lambda} x^\alpha_\lambda = 
  \omega_\alpha^2 x^\alpha_\mu\,.
\end{equation}
Taking only the inside components of the eigenvalue equation, we see
that
\begin{equation}
K'_{a\mu} x^\alpha_\mu = \omega_\alpha^2 x^\alpha_a\,.
\end{equation}
That is to say,
\begin{mathletters}
\begin{eqnarray}
2g x^\alpha_1 - g x^\alpha_2 & = & \omega_\alpha^2 x^\alpha_1 \\
-g x^\alpha_1 + 2g x^\alpha_2 - g x^\alpha_3 & = &
\omega_\alpha^2 x^\alpha_2 \\ 
& \cdots & \nonumber\\
-g x^\alpha_{\Nin-1} + 2 g x^\alpha_\Nin - g x^\alpha_\Ninn
& = & \omega_\alpha^2 x^\alpha_\Nin\,.
\end{eqnarray}%
\end{mathletters}%
Taking $\omega_\alpha$ fixed, there are $\Nin$ equations involving $\Ninn$
unknown components of $\bx^\alpha$.  However, the equations are invariant under
a uniform rescaling of $\bx^\alpha$.  Thus these equations fix
$x^\alpha_\mu$ for $\mu =1\ldots \Ninn$, except for
normalization.  The equations are readily solved, and the solution is
\begin{equation}\label{eqn:definexin}
x^\alpha_\mu = N'_\alpha \sin \mu k'_\alpha\,,
\end{equation}
where
\begin{equation}
\cos k'_\alpha = 1 - {\omega_\alpha^2\over 2 g}\,,
\end{equation}
and $N'_\alpha$ is an unknown normalization factor.
Here $k'_\alpha$ and $N'^\alpha$ can be complex, but
$\xalpha$ must be real.  When $k'_\alpha$ is real we will write $k_\alpha =
k'_\alpha$ and $N_\alpha = N'_\alpha$ and call this a ``normal'' mode.
When $k'_\alpha$ is complex we can write
\begin{mathletters}
\begin{eqnarray}
k'_\alpha & = & \pi + i k_\alpha\\
N'_\alpha & = & i N_\alpha\\
x^\alpha_\mu & = & (-)^{\mu-1} N_\alpha \sinh \mu k_\alpha
\end{eqnarray}%
\end{mathletters}%
with $k_\alpha$ and $N_\alpha$ real.  We will refer to these as
``abnormal'' modes.

As before, we group the eigenvalues into a matrix, $V_{\mu\alpha} =
x^\alpha_\mu$ and arrange the normalizations so that
$K' = {V^{-1}}^T\Lambda V^{-1}$ and $T' = VV^T$.  We know the value of
$T'$ if one of its indices is an inside index,
\begin{equation}\label{eqn:tinin}
\delta_{ab} = T_{ab}  = V_{a\alpha} V_{b\alpha}
\end{equation}
and
\begin{equation}\label{eqn:tinout}
0 = T_{aj} = V_{a\alpha} V_{j\alpha}\,.
\end{equation}
From Eqs.\ (\ref{eqn:definexin}) and (\ref{eqn:tinin}) we get
\begin{equation}\label{eqn:vinin}
\delta_{ab} = \sum_\alpha {N'_\alpha}^2 \sin k'_\alpha a
\sin k'_\alpha b\,.
\end{equation}
Since $V_{\Ninn,\alpha} = N' \sin k'_\alpha (\Ninn)$, we can use 
Eq.\ (\ref{eqn:tinout}) to extend the range of
Eq.\ (\ref{eqn:vinin}) to $1\ldots \Ninn$ except that the equation does not
hold when both $a$ and $b$ have this value.

A similar calculation can be done for $V^{-1}$, the inverse of the
eigenvalue matrix $V$.  In this case we will find that
\begin{equation}
V^{-1}_{\alpha a} = N_\alpha'\sin k'_\alpha a = V_{a\alpha}\,.
\end{equation}
This has the same form as the equation for $V$, but applies only for
$a=1\ldots\Nin$.  That makes $V^{-1}$ less useful than $V$ for
establishing a connection between the inside and the outside region,
and we will not use it further.

\section{Numerical studies}\label{sec:numerical}
We have solved numerically the set of nonlinear equations, Eqs.\
(\ref{eqn:nonlineartosolve}), with
$\beta$ fixed and various values of $\Lin$ and $N$. The problem is one
of solving $\Nout(\Nout+1)$ simultaneous nonlinear equations for
$\Nout(\Nout+1)$ parameters.  In general such problems are quite
difficult to solve, even if we know that there is a unique solution.
Here we used the Powell hybrid method\cite{powell:method}.  If there
are no local minima of the rms error in the function values, this
method converges from any starting point.  Fortunately this appears to
be the case in our problem.  However, Powell's method often converges
quite slowly for large systems, requiring many thousands of iterations
to make progress.  This has limited our numerical solutions to
problems with no more than about 30 oscillators.  The codes were
written in Lisp and executed on DEC Alpha workstations.  The
calculations were done using an arbitrary-precision floating point
package.  The results presented here were computed using at least 38
digits of precision and the solutions found were accurate to at least
21 significant digits.

\begin{figure}
\begin{center}
\leavevmode\epsfbox{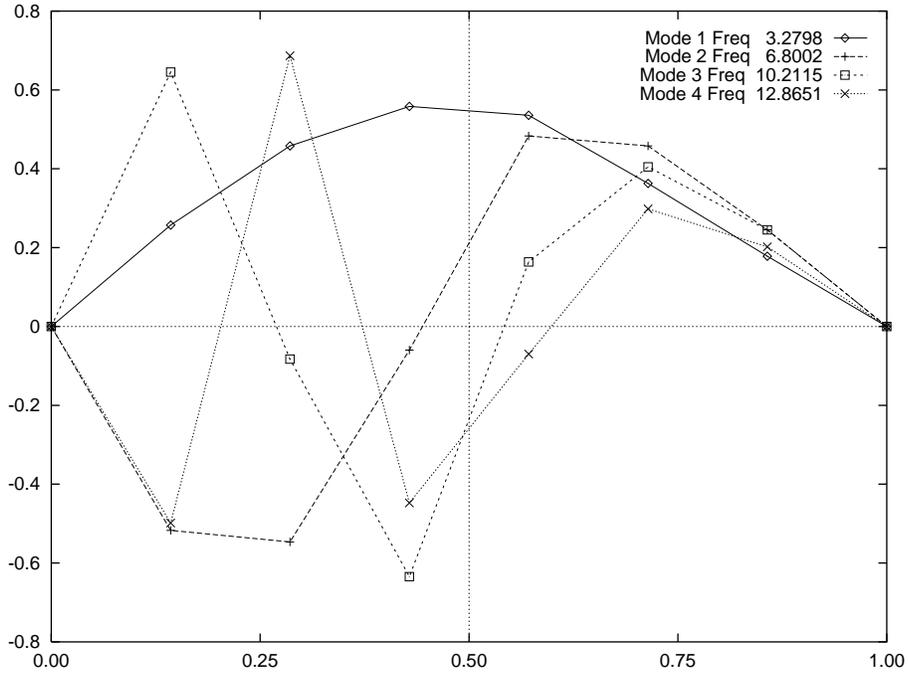}
\end{center}
\caption{Modes and frequencies for the ``normal'' modes of a system
with $L=1.0$, $\Lin=0.5$, $\beta=2$, $N=6$.}
\label{fig:normal-modes}
\end{figure}

\begin{figure}
\begin{center}
\leavevmode
\epsfbox{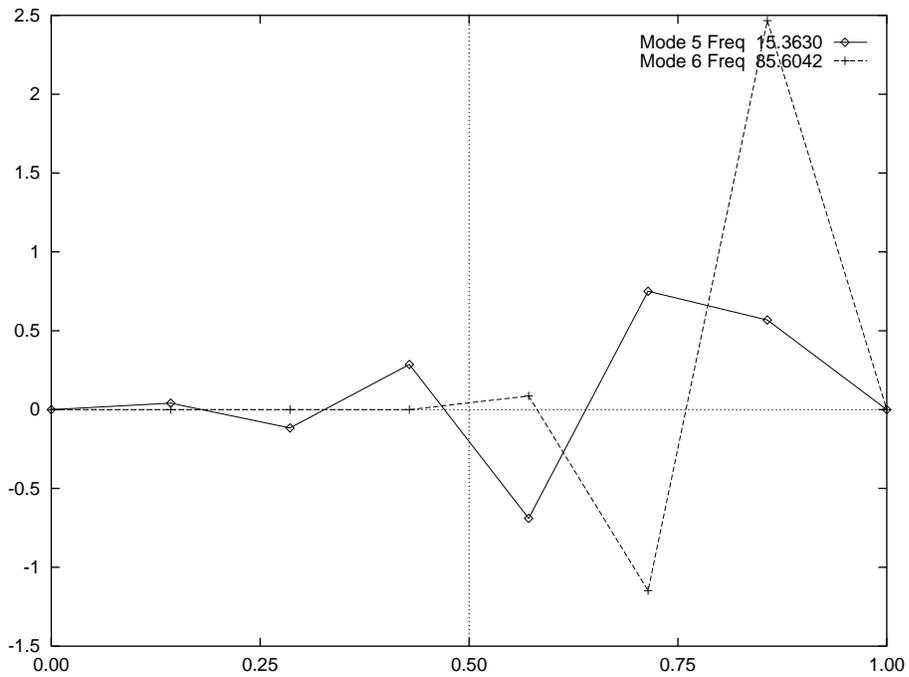}
\end{center}
\caption{Modes and frequencies for the ``abnormal'' modes of a system
with $L=1.0$, $\Lin=0.5$, $\beta=2$, $N=6$.}
\label{fig:abnormal-modes}
\end{figure}

\begin{figure}
\begin{center}
\leavevmode
\epsfbox{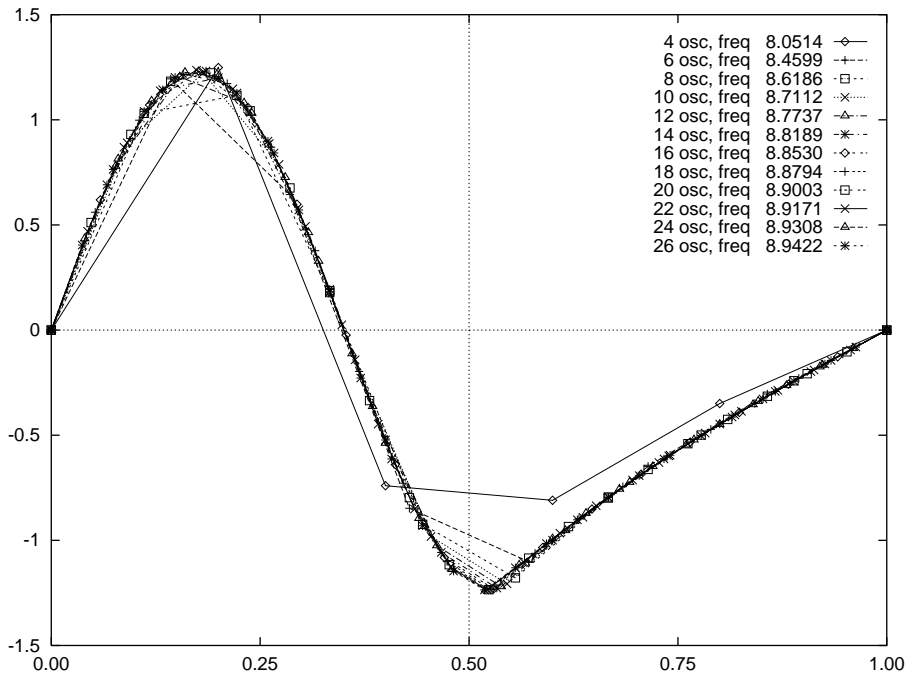}
\end{center}
\caption{The second normal mode, rescaled to the continuum
normalization, for various different number of
oscillators, with $L=1.0$, $\Lin=0.5$, $\beta=0.5$.}
\label{fig:normal-limit}
\end{figure}

\begin{figure}
\begin{center}
\leavevmode
\epsfbox{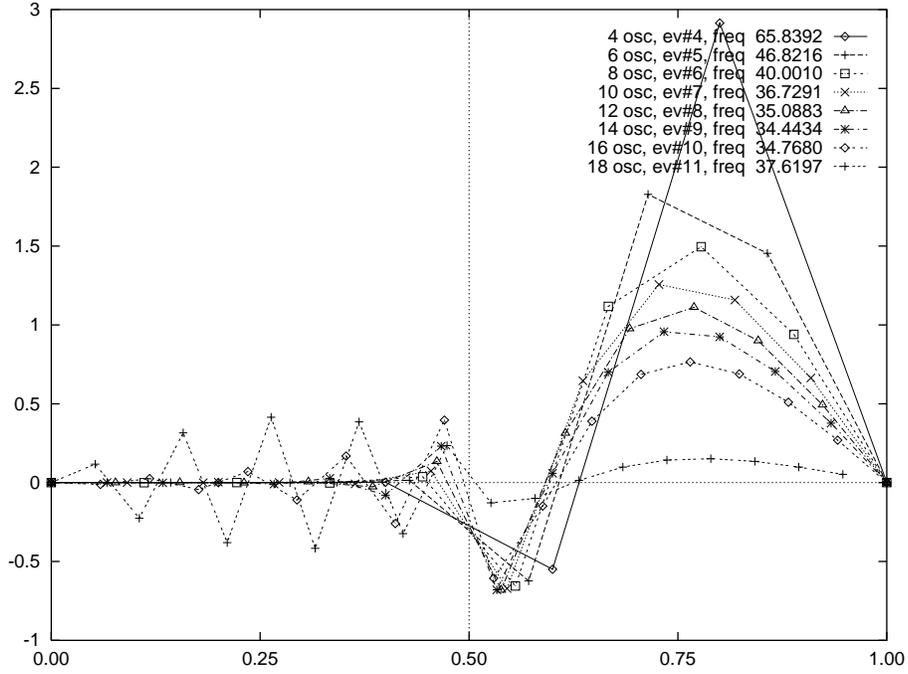}
\end{center}
\caption{An abnormal mode for $L=1.0$, $\Lin=0.5$, $\beta=0.5$.  As $N$ is
increased the mode changes 
smoothly until at $N=18$ there is no corresponding abnormal mode, but
instead one of the new normal modes has a very similar form
but different normalization.}
\label{fig:abnormal-limit}
\end{figure}

To understand the numerical solution we look at the normal mode
frequencies and the forms of the normal modes.  When the mode is
``normal'' (i.e. real $k_\alpha'$ in Eq.\ (\ref{eqn:definexin})) the
mode is a sine wave in the inside region.  When the mode is
``abnormal'' it is essentially a growing exponential.  Typical modes
for a small number of oscillators are shown in Figs.\
\ref{fig:normal-modes} and \ref{fig:abnormal-modes}.
As $N$ becomes large, each ``normal'' mode and its frequency smoothly
approach a limit, providing that we use a normalization appropriate
for the continuum, which means that each mode must be rescaled by
$\sqrt{(N+1)/L}$.  See Fig.\ \ref{fig:normal-limit}.
As $N$ increases, each abnormal mode and its frequency undergo a smooth
evolution, until at some point it 
disappears from the set of abnormal modes and is replaced by a normal
mode \label{sec:modeconvert}
with very similar form in the outside region, as shown in Fig.\
\ref{fig:abnormal-limit}.
Because of this behavior, we believe that if we could solve the
continuum behavior directly we would find just the ``normal'' modes.

\begin{figure}
\begin{center}
\leavevmode
\epsfbox{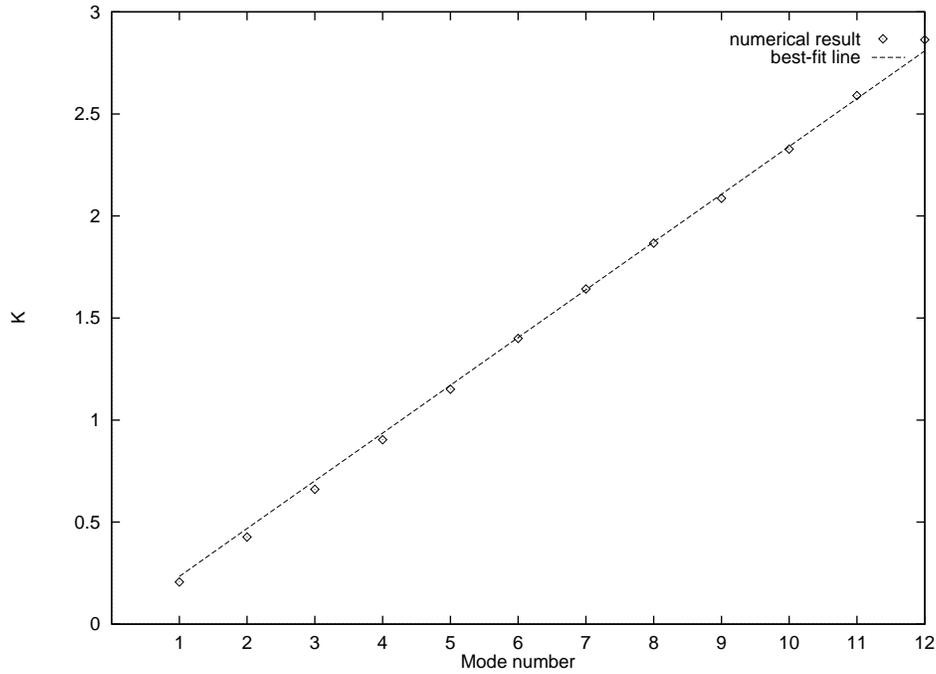}
\end{center}
\caption{The numerically-computed wavenumbers compared with the
best-fit line through the origin for $L=1.0$, $\Lin = 0.5$, $\beta=0.5$,
$N=20$.}
\label{fig:wavenumber-fit}
\end{figure}

The most striking result of the numerical solutions is that the
wavenumbers of the ``normal'' modes are quite evenly spaced.  The
larger the energy, the more accurate is this approximation.  In
Fig.\ \ref{fig:wavenumber-fit},
we show the wavenumbers for $L=1.0$, $\Lin=0.5$, $\beta=0.5$ which
gives an energy of about $0.636$.  Even at this low energy the fit is good to
within a few percent of the typical wavenumber.  For larger energies
the points will lie correspondingly closer to the line.

In the $L\rightarrow\infty$ limit, our problem has only two
dimensionful parameters, $\Lin$ and $E_0$.  Thus there is only one
dimensionless parameter, $\Lin E_0$, that characterizes the problem.
In the 3-dimensional black hole problem the equivalent parameter is $R
E_0$, which for the parameters in Eq.\ (\ref{eqn:goshwownumbers}) is
about $10^{13}$. Thus for application to black holes we are interested
only in very large values of $\Lin E_0$ for which the linear
approximation for the ``normal'' modes will be very good.

Since the ``normal'' mode wavenumbers extend up to $k \sim \pi$, the
number of ``normal'' modes will be given an integer
$\Nnorm \sim \pi/k_1$ where $k_1$ is the interval between wavenumbers.
In addition there will be $N - \Nnorm$ ``abnormal'' modes, with
frequencies $\omega > 2(N+1)/L$.  For $N \gg L/\beta$ these modes do
not contribute to the entropy, because they are exponentially
suppressed.

\section{The first outside oscillator}

In an attempt to minimize our reliance on numerical results we will
use only the statement that the wavenumbers $k_\alpha$ are evenly
spaced.  This leaves one free parameter, the spacing $k_1$, which
depends on the energy $E_0$.  To fix $k_1$ we will look at the
correlator of the position of the first outside oscillator with
itself, which from Eqs.\ (\ref{eqn:xxcorrelator}) and
(\ref{eqn:definexin}) is
\begin{eqnarray}\label{eqn:outoutdef}
\lefteqn{\<x_\Ninn x_\Ninn\>  = 
\sum_\alpha{1\over 2\omega_\alpha}\coth\left({\beta\over 2}\omega_\alpha\right)
\left(x^\alpha_\Ninn\right)^2}\nonumber\\
& & \hspace{20pt}
=  \sum_\alpha{1\over 2\omega_\alpha}\coth\left({\beta\over 2}\omega_\alpha\right)
{N'_\alpha}^2 \sin^2 k'_\alpha(\Ninn)\,.
\end{eqnarray}%
The important point here is that we know $x^\alpha_\mu$ for $\mu$ up
to $\Ninn$, and we know $\<x_\mu x_\nu\>$ for $\mu$ and $\nu$ down to
$\Ninn$.  Thus taking $\mu=\nu=\Ninn$ gives the unique correlator for
which we know the components that go into the expression for the
correlator while also knowing that the correlator must have the same
value as in the vacuum.  The same argument does not work for $\<P_\Ninn
P_\Ninn\>$, because we know $V^{-1}_{\alpha\mu}$ only up to $\mu =
\Nin$ and not $\mu = \Ninn$.

By computing this correlator in the vacuum and in a vacuum-bounded
state for a given $\beta$ we can fix the spacing of the normal mode
frequencies.  In a high energy state, $\beta$ will be
small and so there will be frequencies with $\beta\omega_\alpha \ll
1$.  For such a frequency,
\begin{equation}
\coth {\beta\over 2}\omega_\alpha \approx {2\over\beta\omega_\alpha}
\gg 1\,.
\end{equation}
  In order to keep $\<x_\Ninn^2\>$ the same as in the vacuum
we must require
\begin{equation}
\sin^2 k_\alpha (\Ninn) \ll 1
\end{equation}
 to cancel the large $\coth$,
and consequently
\begin{equation}
k_\alpha (\Ninn) \approx n \pi
\end{equation}
for some integer $n$.  This means that $k_\alpha$ is close to one of the
wavenumbers appropriate to the problem with a rigid boundary at $\Lin$.

\subsection{The vacuum}

First we will compute the correlator for the vacuum, $\<x_\Ninn^2\>_0$.
Here $\beta=\infty$ so the $\coth$ term drops out, giving us
\begin{equation}
\<x_\Ninn^2\>_0 = \sum {1\over2 \omega_n} N_n^2 \sin^2 k_n(\Ninn)\,.
\end{equation}
For the ground state of $N$ oscillators we have
\begin{equation}
N_n^2 = {2\over N+1}\qquad k_n = {n\pi\over N+1}
\end{equation}
and
\begin{equation}
\omega_n = 2{N+1\over L}\sin{k_n\over 2}
\end{equation}
which give
\begin{equation}\label{eqn:xoutoutvacdefined}
\<x_\Ninn^2\>_0 = {L\over 2(N+1)}\sum_{n=1}^N
 {\sin^2 n\pi{\Ninn\over N+1} \over (N+1)\sin {n\pi\over 2(N+1)}}\,.
\end{equation}
We would like to evaluate this expression in the $N\rightarrow\infty$
limit, with $\Nin/N$ fixed.  There is a prefactor $L/(N+1)$, which
goes to zero in this limit, but that is just an artifact of the
conventions we have used for the discrete problem, and will appear in
the finite-energy vacuum-bounded states as well.

We will expand the sum for large N.  There is a logarithmic
divergence.  We are interested in the
$\ln N$ term, and in the constant term, but we will ignore any terms
of order $1/N$ or below.  Accordingly we will take $N+1$ as $N$
wherever it occurs.  

First we expand the numerator using
$\sin^2 x = (1-\cos 2x)/2$.  The first term gives
\begin{equation}
\half\sum_{n=1}^N {1 \over N\sin {n\pi\over 2N}}\,.
\end{equation}
We would like to turn this sum into an integral in the
$N\rightarrow\infty$ limit.  However, we must first subtract out the
divergent part.  Accordingly, we write this as
\begin{equation}
{1\over\pi}\sum_{n=1}^N {1\over n} 
+ \half\sum_{n=1}^N \left({1\over N\sin {n\pi\over 2N}} - {2\over\pi n}
\right)\,.
\end{equation}
The first term gives 
\begin{equation}
{1\over\pi}\left(\ln N + \gamma\right)
\end{equation}
 where $\gamma$ is
Euler's constant.  The second term is finite and can be converted to
an integral in the $N\rightarrow\infty$ limit, to give
\begin{eqnarray}
\half\int_0^1 dx\left({1\over\sin {\pi x\over 2}}-{2\over \pi x}\right)
&=& {1\over\pi}\left[\ln\left({1\over x}\tan{\pi x\over 4}\right)\right]^1_0\\
&=& {1\over\pi}\ln{4\over \pi}\,.
\end{eqnarray}
The remaining term in the sum is
\begin{equation}
-\half\sum_{n=1}^N {\cos{2 n\pi\Nin\over N}\over N \sin{n\pi\over 2 N}}\,.
\end{equation}
To compute this we use
$1/\sin x = \csc x = 1/x + x/6 + 7x^3/360 + \cdots$ to get 
\begin{equation}
-\half\sum_{n=1}^N\cos {2 n\pi \Nin \over N}
\left({2\over\pi n} + {\pi\over 12}{n\over N^2}
+{7\pi\over 2880}{n^3\over N^4} + \cdots\right)\,.
\end{equation}
The first term can be summed,
\begin{equation}
- {1\over\pi}\sum_1^\infty{\cos{2 n\pi \Nin \over N}\over n}
= {1\over\pi}\ln \left(2 \sin {\pi\Nin\over N}\right)\,.
\end{equation}
The rest of the terms do not contribute.  Because of the
oscillations of the cosine, $\sum_1^N n^k \cos (n\pi\Nin/N)$ goes as
$N^k$ rather than $N^{k+1}$ and thus is killed by the corresponding
$N^{k+1}$ in the denominator.

Putting all the terms together we find that

\begin{eqnarray}\label{eqn:vacsumresult}
\<x_\Ninn^2\>_0 = {L\over 2\pi N}\bigg[\ln N & + & \gamma
\ln\left({8\over\pi}\sin{\pi\Nin\over N}\right)\nonumber\\
&+&O\left({1\over N}\right)\bigg]\,.
\end{eqnarray}

\subsection{Using the even-spacing ansatz}\label{sec:bound-calc}

Now we will compute the same correlator in the vacuum-bounded system,
using the ansatz that the wavenumbers are multiples of some spacing
$k_1$, which gives $\Nnorm\approx \pi/k_1$ normal wavenumbers.  This is
equivalent to saying that the normal wavenumbers are those appropriate
to a problem with a rigid boundary at distance $\Lin' =
(\pi/k_1)(L/(N+1)) \approx L(\Nnorm/N)$.  There will also be a
contribution from the abnormal modes.  The contribution from each mode
is positive, so by taking only normal modes we will find a lower bound
on $\<x_\Ninn^2\>$.  Using this bound we can derive an upper bound on
$\Lin'$ and thus on $S(E)$.

We will again work with $N$ oscillators for $N$ large.  We will ignore
all terms of $O(1/N)$ or $O(1/\Nnorm)$.  Accordingly we will take
$\Lin'/\Nnorm = L/N$ and
\begin{equation}
k_n = n k_1 = {\pi n\over \Nnorm} = {\pi n\over \Nnorm+1}\,.
\end{equation}
As before we have
\begin{equation}\label{eqn:defvbfreqs}
\omega_n = {2N\over L}\sin{k_n\over 2} ={2N\over L}\sin{\pi n\over 2\Nnorm}\,.
\end{equation}

Now we can derive the normalizations of the normal modes: We know from
Eqs.\ (\ref{eqn:tinin}) and (\ref{eqn:tinout}) that if $a$ is an
inside oscillator, then for any $\mu$,
\begin{equation}
\sum_\nu V_{a\nu} V_{\mu\nu} = \delta_{a\mu}\,,
\end{equation}
so 
\begin{equation}\label{eqn:orthogonality}
\delta_{a\mu}  = \sum_{n=1}^\Nnorm N_n^2\sin k_n a\sin k_n \mu +
\text{abnormal modes.}
\end{equation}
Later we will find that $\Nnorm$ is close to $\Nin$.  Using this, and
since we are working in the limit where $\Nin$ is large, we can choose
$a$ such that
\begin{equation}\label{eqn:abounds}
1 \ll a \ll 2\Nin - \Nnorm \sim \Nin\,.
\end{equation}
Each abnormal 
mode $\alpha$ contributes 
\begin{equation}
N_\alpha^2 \sinh k_\alpha a \sinh k_\alpha \mu \equiv \delta_{\text{abn}}
\end{equation}
to the right-hand side of Eq.\ (\ref{eqn:orthogonality}).  However, this
same mode contributes
\begin{equation}
{L\over 4 N} N_\alpha^2 {\sinh^2 k_\alpha (\Ninn) \over \cosh (k_\alpha/2)}
\end{equation}
to the sum for $\<x_\Ninn^2\>$ in Eq.\ (\ref{eqn:outoutdef}).
Since $\<x_\Ninn^2\> = L/(2\pi N)[\ln N + O(1)]$ it follows that
\begin{equation}
N_\alpha^2 {\sinh^2 k_\alpha (\Ninn) \over \cosh (k_\alpha/2)} \alt {\ln N\over 2\pi}
\end{equation}
and thus that
\begin{equation}
\delta_{\text{abn}} < {\sinh k_\alpha a\sinh k_\alpha \mu \cosh(k_\alpha/2)\over
	2 \pi \sinh^2 k_\alpha(\Ninn)} \ln N\,.
\end{equation}

The exact values of the $k_\alpha$ for the abnormal modes vary with
$N$.  For successive $N$ values, each $k_\alpha$ decreases toward $0$
until the corresponding mode converts to a normal mode as described in
Sec.\ \ref{sec:modeconvert}.  By avoiding the points where these
``conversions'' are about to take place it is possible to find a
sequence of values for $N$ which have all $k_\alpha \agt 1$.  For such
values,
\begin{equation}
\delta_{\text{abn}} \alt e^{-k_\alpha\left(2 \Nin -a-\mu+3/2\right)}\,.
\end{equation}
From Eq.\ (\ref{eqn:abounds}) the exponent is $\gg 1$, so
$\delta_{\text{abn}}$ is exponentially small, and the contribution of
the abnormal modes to (\ref{eqn:orthogonality}) is suppressed.

Thus we ignore the abnormal modes in Eq.\ (\ref{eqn:orthogonality}),
multiply by $\sin k_m \mu$ and sum over $\mu$ to get
\begin{eqnarray}
\sin k_m a & = & \sum_{\mu=1}^\Nnorm\sin k_m\mu
\sum_n N_n^2\sin k_n a\sin k_n \mu \nonumber\\
& = & \sum_n N_n^2 \sin k_n a \sum_{\mu=1}^\Nnorm
  \sin\text{$\pi m \mu\over \Nnorm+1$}
  \sin \text{$\pi n \mu\over \Nnorm+1$} \nonumber\\
& = & \sum_n N_n^2 \sin k_n a \cdot {\Nnorm+1\over 2} \delta_{mn}\\
& = & N_n^2 {\Nnorm+1\over 2} \sin k_m a\,,
\end{eqnarray}%
from which we conclude that
\begin{equation}
N_n^2 = {2\over\Nnorm+1} \approx {2\over\Nnorm}\,.
\end{equation}
Thus we have
\begin{equation}
\<x_\Ninn^2\> \agt \sum_{n=1}^\Nnorm {\sin^2 k_n\Nin \over\Nnorm \omega_n}
\coth\left({\beta\over 2}\omega_n\right)\,.
\end{equation}
Now define $\Delta$ by $1-\Delta \equiv \Nin/\Nnorm \approx
\Lin/\Lin'$, so that
$\sin^2 k_n \Nin = \sin^2 n\pi (1-\Delta) = \sin^2 n\pi\Delta$.  Then
put in the definition of $\omega_n$ from Eq.\ (\ref{eqn:defvbfreqs}) to get

\begin{eqnarray}\label{eqn:xoutoutdefined}
\<x_\Ninn^2\> \agt {L\over 2N}
&&\sum_{n=1}^\Nnorm\bigg\{
   {\sin^2 n\pi\Delta \over\Nnorm \sin {n\pi\over 2\Nnorm}} \nonumber\\
&& \times \coth\left({\beta N\over L}\sin{n\pi\over 2\Nnorm}\right)\bigg\}\,.
\end{eqnarray}

In the large $N$ limit, the argument of $\coth$ becomes $\beta \pi n
N/ (2 L \Nnorm) \approx \beta\pi n/(2\Lin') = \pi n/(2 \tau')$ where
$\tau'\equiv \Lin'T = \Lin'/\beta$.  Now if it were not for the $\coth$
term, we would have the same sum as in Eq.\
(\ref{eqn:xoutoutvacdefined}) with $\Nnorm$ for $N$ except in the
prefactor. Thus we will exchange $\coth x$ for $1+(\coth x -1) =
1+2/(e^{2x} -1)$ in Eq.\ (\ref{eqn:xoutoutdefined}) to get
\begin{eqnarray}
\<x_\Ninn^2\> \agt {L\over 2\pi N}\bigg[\ln \Nnorm & + & \gamma+
\ln\left({8\over\pi}\sin{\pi\Delta}\right) \nonumber\\
& + & B +O\left({1\over N}\right)\bigg]
\end{eqnarray}
where 
\begin{equation}
B \equiv \pi\sum_{n=1}^\Nnorm {\sin^2 n\pi\Delta \over\Nnorm
\sin {n\pi\over 2\Nnorm}} 
{2\over e^{\pi n / \tau'}-1}\,.
\end{equation}

Now we again use $1/\sin x = 1/x + x/6 + \cdots$.  The first term has
no $\Nnorm$ dependence and we can extend the sum to $\infty$.  In
the next term, the sum is cut off by the exponential in the
denominator, leading to a term of order $(\tau'/\Nnorm)^2$.  Further
terms have higher powers of $\tau'/\Nnorm$.  In the limit
$\Nnorm\rightarrow\infty$ we ignore all these terms, which leaves
\begin{equation}\label{eqn:bsimpsum}
B = 4 \sum_{n=1}^\infty {\sin^2 n\pi\Delta \over
n \left(e^{\pi n/\tau'}-1\right)}\,.
\end{equation}
We are interested in the high-energy limit, for which $\tau' \gg 1$.
Later we will see that $\Delta$ is of order $\ln\tau'/\tau' \ll 1$.
Thus the summand in Eq.\ (\ref{eqn:bsimpsum}) is slowly varying and we
can convert the sum into an integral,
\begin{equation}\label{eqn:bint}
B \approx 4\int_0^\infty {\sin^2 \pi\Delta x\, dx\over
x\left(e^{\pi x/\tau'}-1\right)}\,.
\end{equation}
The error in Eq.\ (\ref{eqn:bint}) is approximately the term that we
would have for $n=0$ in Eq.\ (\ref{eqn:bsimpsum}).  Taking the
$n\rightarrow 0$ limit we find that this term has order $\Delta^2\tau'
\sim (\ln\tau')^2/\tau' \ll 1$, so our approximation is good.
The integral in Eq.\ (\ref{eqn:bint}) can be done and the result
is
\begin{equation}\label{eqn:bfinal}
B = 2 \pi\tau'\Delta + \ln{1-e^{-4 \pi\tau' \Delta}\over 4 \pi\tau' \Delta}\,.
\end{equation}

Now we set $\<x_\Ninn^2\>$ = $\<x_\Ninn^2\>_0$ to get
\begin{eqnarray}
\lefteqn{\ln \Nnorm+\gamma+\ln\left({8\over\pi}\sin{\pi\Delta}\right) + B}
     \nonumber \\
&& \hspace{20pt} \le \ln N+\gamma+
\ln\left({8\over\pi}\sin{\pi\Nin\over N}\right)
\end{eqnarray}
or
\begin{equation}
B \le \ln\left({N\over \Nnorm}{\sin{\pi\Nin\over N}\over\sin\pi\Delta}\right)
= \ln\left({L\over\Lin'}{\sin{\pi\Lin\over L}\over\sin\pi\Delta}\right)\,.
\end{equation}
Note that we have taken the limit $N\rightarrow\infty$ and there is no
longer any dependence on $N$.  
If we now let $L\rightarrow\infty$ with
$\Lin$ fixed, we can approximate $\sin(\pi\Lin/L) = \pi\Lin/L$ to get
\begin{equation}
B \le \ln {\Lin\over\Lin' \sin\pi\Delta} = \ln {1-\Delta\over\sin\pi\Delta}
\approx \ln {1\over\pi\Delta}
\end{equation}
since $\Delta$ is small.

Using $B$ from Eq.\ (\ref{eqn:bfinal}) we get
\begin{equation}
2\pi\tau'\Delta + \ln{1-e^{-4 \pi\tau' \Delta}\over 4 \pi\tau' \Delta}
\le \ln{1\over\pi \Delta}\,.
\end{equation}
Thus $\Delta \le \Delta_{\max}$ where
\begin{equation}\label{eqn:deltamax}
\Delta_{\max}
 = {1\over 2 \pi\tau'}\ln{4\tau'\over 1-e^{-4\pi\tau'\Delta_{\max}}}\,.
\end{equation}
If instead of $\tau' = \Lin' T$ we use $\tau \equiv \Lin T$ we will make an
error of order $\Delta_{\max}$, which we expect to be small.  So we ignore
the second order contribution and take $\tau'$ as $\tau$ in
Eq.\ (\ref{eqn:deltamax}),
\begin{equation}
\Delta_{\max}
 = {1\over 2 \pi\tau}\ln{4\tau\over 1-e^{-4\pi\tau\Delta_{\max}}}\,.
\end{equation}
If we ignore $e^{-4\pi\tau\Delta_{\max}}$ in the denominator we get
\begin{equation}
\Delta_{\max} = {1\over 2\pi\tau}\ln{4\tau}\,.
\end{equation}
Using this we find that 
$e^{-4\pi\tau\Delta_{\max}} = (4\tau)^{-2} \ll 1$
since $\tau \gg 1$, which justifies ignoring this term.  We will also
ignore $\ln 4$ by comparison with $\ln \tau$.
Thus we conclude
\begin{equation}
\Delta \alt {1\over 2\pi\tau}\ln \tau  + O\left({1\over\tau}\right)
= {1\over 2\pi\Lin T}\ln{\Lin T} + O\left({1\over\Lin T}\right)
\end{equation}
and
\begin{equation}
\Lin' \le \Lin + {1\over 2\pi T} \ln {\Lin T} +O\left({1\over T}\right)\,.
\end{equation}
The equivalent system is larger by at most a thermal wavelength times a
logarithmic factor depending on the inside size.

\section{Propagation of bounds}\label{sec:bound-prop}

In the previous section we derived an expression that gives the
frequencies, and thus the entropy, for a vacuum-bounded system at a
given temperature $T = 1/\beta$.  Given such an expression, we would
like to compute the entropy as a function of energy.  But the energy
is not simple to compute from the frequencies alone.\footnote{It can
be done, but since $E$ needs to be renormalized against the
ground-state energy of the entire system the result depends
sensitively on the frequencies and normalizations even for very
high-energy modes.}  However, we can easily compare the entropy of the
vacuum-bounded system to that of a system with a rigid boundary at
$\Lin$ and the same temperature.  To make this comparison at fixed
energy instead, we proceed as follows.

Consider the free energy $F=E-TS$ which has $dF= -S dT$.  Integrating
gives
\begin{equation}\label{eqn:fdef}
E-TS= -\int_0^T S(T') dT'\,.
\end{equation}%
Let $S\rb$, $T\rb$ denote the entropy and temperature of the system
with a rigid boundary at $\Lin$, and $S$, $T$, and so on denote those
for the vacuum bounded system. For any quantity $A$ let $\delta A(T)$
denote the difference between vacuum-bounded and rigid-bounded systems
at fixed temperature, $\delta A(T) \equiv A(T) - A\rb(T)$, and $\delta
A(E)$ denote the same difference at fixed energy, $\delta A(E) \equiv
A(E) - A\rb(E)$.  With $E$ fixed we compare the differences (to first
order) in the two sides of Eq.\ (\ref{eqn:fdef}) between the
vacuum-bounded and rigid-bounded systems,
\begin{equation}
-T \delta S(E) - \delta T(E) S = - \delta T(E) S - \int_0^T \delta
S(T') dT'\,,
\end{equation}
where the first term on the right-hand side comes from the change in
the integration limit.  Thus
\begin{equation}
\delta S(E) = {1\over T} \int_0^T \delta S(T') dT'\,.
\end{equation}

\section{Deriving the bound}

Now we apply these results to the case from Sec.\ \ref{sec:bound-calc}
where
\begin{equation}
\omega_n \approx {n\pi\over \Lin'}
\qquad\text{and}\qquad
\Lin' \approx \Lin(1+\Delta)
\end{equation}
with
\begin{equation}
\Delta \Lin \le {1\over 2\pi T}\ln \Lin T\,.
\end{equation}
This says that at any given temperature, the vacuum-bounded system has
the entropy $S(T)$ of a system of length $\Lin'$.  Now in a
one-dimensional system the entropy density is proportional to the
temperature, 
\begin{equation}
S\rb =  {\pi\over 3}\Lin T\,,
\end{equation}
and thus the entropy difference between vacuum-bounded and
rigid-bounded systems is
\begin{equation}
\delta S(T) = {\pi\over 3}\Delta\Lin T \le {1\over 6}\ln\Lin T\,.
\end{equation}
Using the results from Sec.\ \ref{sec:bound-prop},
\begin{equation}
\delta S(E) \le {1\over T}\int_0^T {1\over 6}\ln\Lin T' dT'
= {1\over 6}\left(\ln\Lin T-1\right)\,.
\end{equation}
Since we are ignoring terms of order $1$ by comparison with those
of order $\ln T$, we can write this\footnote{It has happened that
$\delta S(E)$ and $\delta S(T)$ are approximately the 
same, but that is a particular property of the system at hand.  For
example, if $\Delta$ were a constant we would have
$\delta S(E) = 1/T\int_0^Tc\Delta\Lin T' dT' = 2 c\Delta\Lin T
= 2 \delta S(T).$}
\begin{equation}
\delta S(E) \alt {1\over 6}\ln\Lin T\,.
\end{equation}

So, we conclude that the vacuum-bounded condition closely approximates
the rigid box of length $\Lin$.  For the same energy, the
vacuum-bounded condition allows slightly more entropy.  The entropy
difference grows at most logarithmically with the number of thermal
wavelengths in the inside region.  Since the total entropy at a given
temperature is linear in $T$, the ratio of $\delta S$ to $S$ goes to
zero at high temperatures.

\section{Discussion}
We have introduced a new way of specifying that matter and energy are
confined to a particular region of space.  Rather than giving a
boundary condition per se, we specify a condition on a density matrix
describing the state of the overall system.  We require that any
measurement which does not look into the inside region cannot
distinguish our system from the vacuum.  This avoids certain
difficulties such as the Casimir energy that results from the
introduction of a boundary and the geometric entropy
\cite{srednicki:geom-ent,callan:geom-ent} that results from ignoring
part of a system.  For these ``vacuum-bounded'' states, we consider
the problem of finding the maximum-entropy state for a given total
energy.  This is analogous to the problem of finding the thermal
state in a system with a rigid boundary.

Unfortunately, the vacuum-bounded problem is more difficult than
the analogous problem with a rigid boundary and we must resort to
working in one dimension and to numerical solution on a lattice.  From
the numerical solution we justify the ansatz that the continuum
wavenumbers are evenly spaced in this problem.  Using this ansatz we
compute an upper bound on the entropy of a vacuum-bounded state, and
show that for high energies ($ER \gg 1$) the entropy approaches that
of a system with rigid boundaries.  Of course this is what one would
expect for a system whose typical wavelengths are much shorter than
the size of the inside region.

To apply this result to an evaporating black hole we look at the state
produced by the black hole after evaporation \cite{preskill:review}.
Since our calculation was one-dimensional we must assume that the
similarity between the vacuum-bounded state and the thermal state with
a rigid boundary extends to 3 dimensions.  Then we infer that very
little entropy can be emitted in the final explosion, confirming the
results of Aharonov, Casher and Nussinov \cite{aharonov:orig} and
Preskill \cite{preskill:review}.  For example, a black-hole formed in
the big bang with mass of order $10^{15}g$ would be evaporating today.
During its life it would have radiated entropy $S\sim 10^{38}$.  Now
we assume that the entropy of the final explosion has energy $E\sim
10^{19}\text{erg}$ contained in radius $R\sim 10^{-23}\text{cm}$ as in Sec.\
\ref{sec:realistic-er}, and that the maximum entropy is not too
different from that of a spherical box, in accord with our
1-dimensional result.  Then we find that the final explosion can emit
only entropy $S \sim 10^{10}$, which is a factor of $10^{28}$ too
little to produce a pure state.

This argument means that either a black hole must not evaporate
completely but rather leave a remnant or remnants, that information
must be lost, or else that the Hawking radiation is not exactly
thermal, even at very early times \cite{esko:nosemiclass}.

\section{Acknowledgments}
We would like to thank Alan Guth and Sean Carroll for much advice and
assistance; Niklas Dellby, Eddie Farhi, Dan Freedman, Ken Halpern, Andy Latto,
Arthur Lue, Keith Ramsay and Peter Unrau for helpful conversations;
Harlequin Inc.\ for providing a copy of their LispWorks product on
which some of the computations were done; Bruno Haible and Marcus
Daniels for their CLISP Common Lisp implementation; and Kevin
Broughan and William Press for making available a version of {\em Numerical
Recipes} translated into LISP\footnote{These routines and many other
 are now available on a Numerical Recipes CDROM\protect\cite{numrecip:cdrom}}.

This work is supported in part by funds provided by the
U.S. Department of Energy (D.O.E.) under cooperative research
agreement DE-FC02-94ER40818.

\appendix
\section{Concavity of the Entropy}\label{app:convex}

Lieb \cite{lieb:convex} showed that $S$ is always (downward)
concave.  Here we obtain the same result by a different technique and
show also that the concavity is strict, i.e.\ that $d^2S/dt^2$ never
vanishes.

We consider $S = S(\rho') = - Tr \rho' \ln \rho'$ with $\rho' \equiv
\rho+t\delta\rho$ for some $\delta\rho$ with $\Tr\delta\rho = 0$.  Then
\begin{equation}
{dS\over dt} = - \Tr \delta\rho \ln \rho'
\end{equation}
and
\begin{equation}
{d^2S \over dt^2} = - \Tr \delta\rho {d\over dt}\ln\rho'\,.
\end{equation}
To expand this we use the formula
\begin{equation}\label{eqn:dln}
{d\over dt}\ln A = \int_0^1 ds \left(I-s(A-I)\right)^{-1} {dA\over dt}
\left(I-s(A-I)\right)^{-1}
\end{equation}
which can easily be derived as follows: Let $B = A - I$.  We expand
\begin{equation}
\ln A = \ln(I+B) = B + {B^2\over 2} + {B^3\over 3} + \cdots
\end{equation}
and differentiate to get
\begin{eqnarray}
\lefteqn{{d\over dt} \ln A = {dB\over dt}
{1\over 2}\left({dB\over dt}B + B{dB\over dt}\right)}\hspace{20pt}&\nonumber\\
&& + {1\over 3}\left({dB\over dt}B^2 + B{dB\over dt}B + B^2 {dB\over dt}\right)
+ \cdots\,.
\end{eqnarray}
We observe that this is
\widetext
\begin{eqnarray}
\lefteqn{\int_0^1 ds {dB\over dt} +
s\left({dB\over dt}B + B{dB\over dt}\right)
 + s^2\left({dB\over dt}B^2 + B{dB\over dt}B +
B^2 {dB\over dt}\right) }\hspace{10pt} & \nonumber\\
& = & \int_0^1 ds \left(1+sB+s^2B^2+\cdots\right) {dB\over dt}
     \left(1+sB+s^2B^2+\cdots\right) \nonumber\\
& = & \int_0^1 ds \left(I-sB\right)^{-1} {dB\over dt} 
         \left(I-sB\right)^{-1}
\end{eqnarray}
\narrowtext
as desired.

Using Eq.\ (\ref{eqn:dln}) we find
\begin{eqnarray}
{d^2S\over dt^2} &=& - \int_0^1 ds \Tr \delta\rho (I-s(\rho-I))^{-1}
   \delta\rho \left(I-s(\rho-I)\right)^{-1}\nonumber\\
&=& - \int_0^1 \Tr X(s)^2
\end{eqnarray}
where
\begin{equation}
X(s) \equiv (\delta\rho)^{1/2} (I-s(\rho-I))^{-1} (\delta\rho)^{1/2}\,.
\end{equation}
Since $X(s)$ is Hermitian, $\Tr X(s)^2 \ge 0$ with equality obtained
only for $X(s) = 0$.  Thus
\begin{equation}
{dS^2\over dt^2} \le 0
\end{equation}
with equality only if $X(0) = 0$ for all $s$.  But $X(0) =
\delta\rho$, and thus
\begin{equation}
{dS^2\over dt^2} < 0
\end{equation}
for any nonzero $\delta\rho$.


\begin{thebibliography}{10}

\bibitem{hawking:orig}
S. Hawking, Phys. Rev. D {\bf 248},  30  (1974).

\bibitem{page:review}
D.~N. Page,  in {\em Proceedings of the 5th Canadian Conference on General
  Relativity and Relativistic Astrophysics}, edited by R. Mann and R.
  McLenaghan (World Scientific, River Edge, NJ, 1994), pp.\ 1--41.

\bibitem{preskill:review}
J. Preskill,  in {\em International Symposium on Black Holes, Membranes,
  Wormholes, and Superstrings} (World Scientific, River Edge, NJ, 1993), pp.\
  22--39.

\bibitem{banks:review}
T. Banks,  in {\em String Theory, Gauge Theory, and Quantum Gravity}, edited by
  R. Dijgraaf, K. Narain, and S. Randjbar-Daemi (North-Holland, Amsterdam,
  1995), pp.\ 21--65.

\bibitem{hawking:virtual}
S.~W. Hawking, Phys. Rev. D {\bf 53},  3099  (1996).

\bibitem{giddings:infoloss}
S.~B. Giddings, Phys. Rev. D {\bf 49},  4078  (1994).

\bibitem{banks:pure-mixed}
T. Banks, L. Susskind, and M.~E. Peskin, Nucl. Phys. {\bf B244},  125  (1984).

\bibitem{strominger:unitary}
A. Strominger, preprint USCSBTH-94-34, {\em Unitary Rules for Black Hole
  Evaporation}, hep-th/9410187.

\bibitem{strominger:baby}
A. Strominger and J. Polchinski, Phys. Rev. D {\bf 50},  7403  (1994).

\bibitem{susskind:strings}
L. Susskind and J. Uglum,  in {\em String theory, gauge theory and quantum
  gravity: proceedings of the Trieste Spring School and Workshop, ICTP,
  Trieste, Italy, 27 March-7 April, 1995}, edited by R. Dijkgraaf
  (North-Holland, Amsterdam, 1996).

\bibitem{verlinde:complement}
Y. Kiem, H. Verlinde, and E. Verlinde, Phys. Rev. D {\bf 52},  7053  (1995).

\bibitem{esko:nosemiclass}
E. Keski-Vakkuri, G. Lifschytz, S.~D. Mathur, and M.~E. Ortiz, Phys. Rev. D
  {\bf 51},  1764  (1995).

\bibitem{bose:nosemiclass}
S. Bose, L. Parker, and Y. Peleg, Phys. Rev. D {\bf 54},  no. 10  (1996), to
  appear.

\bibitem{casher:nosemiclass}
A. Casher {\it et~al.}, preprint TAUP-2344-96, {\em Black hole horizon
  fluctuations}, hep-th/9606106.

\bibitem{strominger:string}
A. Strominger and C. Vafa, Phys. Rev. Lett. {\bf B379},  99  (1996).

\bibitem{esko:string}
E. Keski-Vakkuri and P. Kraus, preprint CALT-68-2079, hep-th/9610045, and
  references therein.

\bibitem{aharonov:orig}
Y. Aharonov, A. Casher, and S. Nussinov, Phys. Lett. B {\bf 191},  51  (1987).

\bibitem{wilczek:mirror}
F. Wilczek,  in {\em International Symposium on Black Holes, Membranes,
  Wormholes, and Superstrings} (World Scientific, River Edge, NJ, 1993), pp.\
  1--21.

\bibitem{bekenstein:talk}
J.~D. Bekenstein,  in {\em General Relativity: proceedings of the 7th Marcel
  Grossman Meeting}, edited by R. Ruffini and M. Keiser (World Scientific,
  River Edge, NJ, 1995).

\bibitem{bekenstein:argue}
J.~D. Bekenstein, Phys. Rev. D {\bf 27},  2262  (1983).

\bibitem{unruh-wald:argue}
W.~G. Unruh and R.~M. Wald, Phys. Rev. D {\bf 27},  2271  (1983).

\bibitem{page:radiation}
D.~N. Page, Phys. Rev. D {\bf 13},  198  (1976).

\bibitem{srednicki:geom-ent}
M. Srednicki, Phys. Rev. Lett. {\bf 71},  666  (1993).

\bibitem{callan:geom-ent}
C. Callan and F. Wilczek, Phys. Lett. B {\bf 333},  55  (1994).

\bibitem{lieb:convex}
E.~H. Lieb, Bull. Am. Math. Soc. {\bf 81},  1  (1975).

\bibitem{powell:method}
M.~J.~D. Powell,  in {\em Numerical Methods for Nonlinear Algebraic Equations},
  edited by P. Rabinowitz (Gordon and Breach, London, New York, 1969), Chap.~7,
  pp.\ 87--114.

\bibitem{numrecip:cdrom}
W.~H. Press, S.~A. Teukolsky, W.~T. Vetterling, and B.~P. Flannery, {\em
  Numerical Recipes: The Art of Scientific Computing, Code CDROM v 2.06}
  (Cambridge University Press, Cambridge, 1996).

\end{thebibliography}
\end{document}